\documentclass{revtex4-1}

\usepackage{graphicx,psfrag}
\usepackage{epstopdf}
\usepackage{amsmath}
\usepackage{bm}
\usepackage{amssymb}
\usepackage{subfigure}

\newcommand{\pdone}[2]{{\frac{\partial #1}{\partial #2}}}

\begin{document}

\title{\large \bf Two-Dimensional Structure-Embedded Acoustic Lenses based on\\ Periodic Acoustic Black Holes}

\author{Hongfei Zhu}
\email{hongfei.zhu.44@nd.edu}
\affiliation{Department of Aerospace and Mechanical Engineering, University of Notre Dame, Notre Dame, IN 46556}
\author {Fabio Semperlotti}
\email{fsemperl@purdue.edu}
\affiliation{School of Mechanical Engineering, Purdue University, West Lafayette, IN 47907}

\begin{abstract}
Recent studies have introduced a new class of two-dimensional acoustic metamaterials whose dispersion and propagation properties results from the use of geometric inhomogeneities in the form of Acoustic Black Holes (ABH). The ABH is an element able to smoothly bend and slow down elastic waves, therefore providing a variety of unconventional dispersion and propagation properties typically observed in more complex multi-material and locally resonant designs. This approach enables thin-walled structural elements having fully embedded acoustic lenses capable of different functionalities such as focusing, collimation, and negative refraction. Numerical simulations show that such structures exhibit broadband operating conditions that span both the metamaterial and the phononic range. Full field experimental measurements allow validating the design approach and confirm the performance of the embedded acoustic lens.
\end{abstract}

\maketitle
\section{Introduction}
Focusing \cite{focusing1,focusing2,focusing3,focusing4,focusing5,focusing6,focusing7}, collimation \cite{collimation1,collimation2,collimation3,collimation4}, sub-wavelength resolution \cite{sub1,sub2,sub3,sub4}, and negative refraction \cite{Pendry} have been among the most fascinating properties observed in either electromagnetic or acoustic metamaterials. Within the range of validity of the long wavelength approximation, these artificial media are often modeled using effective parameters obtained via a homogenization approach\cite{EM1,EM2}. Several studies\cite{NP1,NP2,Liu} have also shown that the effective material properties can be made either zero or negative by exploiting the effect of local resonances.

In the acoustic field, the use of metamaterials for practical applications has been severely limited by their fabrication complexity. To-date, the majority of the locally-resonant designs exploits a multi-material approach that relies on metal inclusions embedded inside a polymeric host that serves as coupling with the main material. This design approach is affected by fabrication issues (particularly those due to interfacing highly dissimilar materials) and does not yield structural (i.e. load-bearing) materials.

Recently, Zhu et al. \cite{Zhu3} have introduced a new class of two-dimensional single-material systems able to achieve high level dispersion and propagation properties comparable with those of traditional multi-material designs \cite{Liu} while also combining features proper of both non-resonant and locally-resonant systems\cite{cui}. Zhu's design focused on a thin two dimensional waveguide tailored via an embedded lattice of geometric tapers characterized by a power-law thickness variation; an element known as the Acoustic Black Hole (ABH) (Fig. \ref{Fig1}a). Among the distinctive features of the ABH, we highlight the progressive and smooth reduction of the phase velocity as flexural waves approach its center. Zhu's design exploits geometric inhomogeneities, fabricated in an initially homogeneous and isotropic thin plate, to tailor the dynamic properties of the host structure. This approach drastically reduces the fabrication complexity while increasing the scalability of the design because the inhomogeneities can be easily obtained by machining prescribed taper profiles. We merely observe that this procedure could virtually transform any material into a metamaterial exhibiting carefully controlled wave propagation properties. It is worth noting that the tapered design does not alter the structural character of the original component which maintains its load-bearing capabilities. This characteristic is in net contrast with multi-material designs in which the interfaces between dissimilar materials (typically metals and polymers) drastically affect fabrication complexity, structural performance, and durability.

While previous work [\onlinecite{Zhu3}] concentrated on the analysis and understanding of the dispersion properties in connection with the specific ABH lattice structure and its design parameters, this study investigates the potential of the ABH-based design to synthesize devices able to manipulate acousto-elastic waves in structural waveguides. In particular, this paper explores the use of tapered units to design and implement different types of acoustic lenses fully embedded in structural waveguides and able to achieve unconventional propagation effects such as focusing, collimation, and bi-refraction.

Acoustic engineered materials are typically designed for specific (wavelength) operating ranges which are governed by different propagation mechanisms. Metamaterials are designed for the long wavelength limit and are governed by effective material properties obtained according to homogenization principles. Phononic materials are designed for the short wavelength limit and are governed by scattering mechanisms. The two designs drastically under perform when used outside their respective operating range. It is anticipated that the ABH-based design exploits the same operating principle in both the long and the short wavelength limits, therefore providing a design approach with an intrinsic potential for broadband performance.

 As previously mentioned, the unit cell at the basis of the metamaterial design exploits a type of tapered profile known as the \textit{Acoustic Black Hole} \cite{Krylov1,Krylov2}. The ABH is a structural feature, typically embedded in plate-like components, often used for the control of vibration \cite{krylovadd,krylovadd1,krylovadd2,Zhao,Colon1,Colon2} and sound radiated by structural elements\cite{Colon,Colon3}. The ABH is obtained by embedding a power-law taper having a thickness profile of the form $h(x)=\varepsilon x^m$.

The ABH determines a spatially dependent distribution of the phase and group velocity. In a one-dimensional power-law taper, the phase $c_p$ and group velocity $c_g$\cite{Mironov} are:
\begin{subequations}
\begin{gather}
              c_p(x)=\sqrt[4]{\frac{D}{\rho h^3}}\sqrt{\omega \cdot \varepsilon x^m} \label{velocityp} \\
              c_g(x)=\sqrt[4]{\frac{16D}{\rho h^3}}\sqrt{\omega \cdot \varepsilon x^m} \label{velocityg}
\end{gather}
\end{subequations}
where $D(x)$ is the plate bending stiffness. The wavenumber becomes a function of the position along the taper:
\begin{equation}
k(x)=\sqrt[4]{12} \sqrt{\frac{k_{l}}{\varepsilon x_{m}}} \label{localk}
\end{equation}
where $k_{l}$ is the longitudinal wavenumber in a thin plate of constant thickness $h_0=h(x_0)$, with $x_0$ being an arbitrary position for the calculation of the local thickness.

Extending these concepts to an axis-symmetric geometry, the wavevector becomes a spatial function $k(r,\theta)$, where $r$ and $\theta$ are in-plane cylindrical coordinates. The trajectory of an acoustic wave traveling through the ABH would be bent in the direction of the negative phase velocity gradient, i.e. $\pdone{c_p}{r} \leq 0$, which coincides with any radial path in the inward direction (that is the direction of the decreasing radius). When the smoothness condition\cite{Mironov} $\{ m$, $\varepsilon \}$ $\in \mathbb{R}$, $m \geq 2$, and $\varepsilon \ll (3 \rho \omega^2 / E)^{1/2}$ is satisfied, reflections are minimized and the wave is eventually trapped inside the ABH. Note that this condition would be achieved only in presence of an ideal ABH profile for which the thickness decreases to zero at the center of the taper. In real structures, the residual thickness at the ABH center $h(r)=\epsilon r^m+h_r$ is non-zero as shown in Fig.\ref{Fig1}a.  This condition results in selected frequencies (those associated with the higher phase velocities) being able to cross entirely the tapered area although with a curved trajectory.

In the following study, we present the application of two-dimensional ABH periodic lattices to the design of acoustic lenses fully embedded into structural waveguides. Unconventional propagation effects including focusing, collimation, and bi-refraction were numerically investigated for both the long and the short wavelength operating range. A subset of these lenses was also fabricated and experimentally validated.

\section{Embedded ABH acoustic lens: numerical model}

\begin{figure}[h!]
 \center{\includegraphics[scale=0.42]{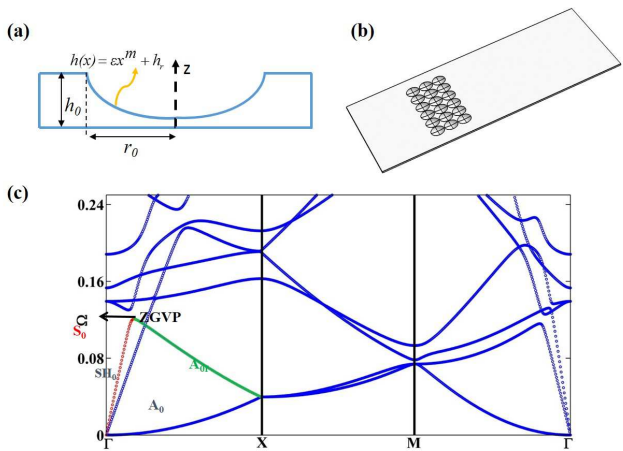}}
  \caption{(a) Schematic of the waveguide with embedded periodic arrays of Acoustic Black Holes and (b) cross-section of an ABH. (c) Band structure of the studied acoustic metamaterials made of periodic geometric tapers for the frequency range up to $\Omega$=0.25.} \label{Fig1}
\end{figure}

For the present study, we embedded a finite-size slab of a squared periodic lattice in an otherwise flat plate. The corresponding numerical model is shown in Fig.\ref{Fig1}a. The waveguide thickness was $h_0=8 mm$ while the ABH taper was defined by the coefficients $m=2.2$ and $h_r=1.1 mm $. The radius of the ABH was $r_0=50 mm$ while the lattice constant was $a=140 mm$. The waveguide can be conceptually divided in three parts one of which is the periodic slab while the remaining two are the constant-thickness layers before and after the slab (Fig. \ref{Fig1}a). The ABH slab represents the acoustic lens while the homogeneous layers are used to generate and receive the incident and the transmitted wave. The numerical model was solved by finite element analysis using the commercial software COMSOL Multiphysics. Perfectly Matched Layers (PML) were used on the surrounding boundaries to minimize the effect of reflections. A frequency analysis was performed to extract the propagation behavior.

Note that the dispersion properties of a thin plate with an embedded periodic lattice of ABHs were studied in depth in \cite{Zhu3} and they will only be briefly summarized in the following section in order to facilitate the interpretation of the numerical and experimental results.

\section{LONG WAVELENGTH LIMIT: BAND STRUCTURE AND PROPAGATION ANALYSIS}

In this section we investigate, both numerically and experimentally, the performance of the ABH-based embedded acoustic lens in the long wavelength limit, that is when the wavelength of the flexural wave in the flat structure is, at least, twice the size of the unit cell. This is often referred to as the \textit{metamaterial range}. In this range, the fundamental propagation characteristics of the unit cell can be well described by the band structure and the Equi-Frequency-Contours (EFC) \cite{Zhu3}. We highlight that among the most noticeable properties of the ABH plates mode hybridization, bi-refraction, and Dirac-like cones were observed.

\begin{figure}[h!]
 \center{\includegraphics[scale=0.45]{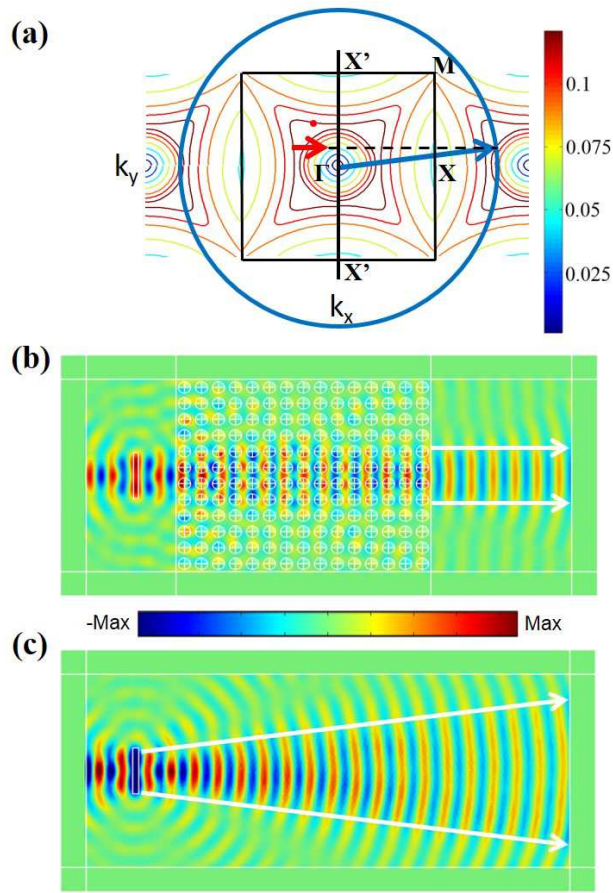}}
  \caption{(a) The equi-frequency-contours of the hybrid $S_0-A_{0f}$. The superimposed schematics shows the Brillouin zone (black line) and the $k-$conservation analysis that allows identifying the direction of the refracted wave. (b) The out-of-plane displacement field in the presence of the acoustic lens showing the strong collimation effect. (c) The out-of-plane displacement field generated by an excitation at $f_0=2.2$ kHz in the absence of the acoustic lens showing the expected diverging ultrasonic beam.} \label{Fig2}
\end{figure}

The dispersion curves of the ABH waveguide are shown in Fig.\ref{Fig1}c in terms of the normalized frequency. We focus on the first folded $A_0$ mode (marked green in Fig.\ref{Fig1}c), because this is the fundamental flexural mode which is largely excited in structural waveguides and also because of a peculiar hybridization effect associated with positive-negative group velocity. The equi-frequency-contours of this mode are presented in Fig.\ref{Fig2}a. A quick inspection of the EFCs reveals that a large number of different propagation behaviors should be expected in this lattice. Moving from the $\Gamma$ point towards increasing wavenumbers, say along the $\Gamma - M$ direction, we observe circular, square, concave, and convex EFC, respectively. While the circular profile concerns an isotropic propagation typical of very long wavelengths, the square and concave EFCs suggest collimation and focusing effects, respectively. The convex EFCs were shown numerically \cite{Zhu3} to correspond to bi-refraction behavior.

For the numerical analysis of the bi-refraction effect the reader is remanded to [\onlinecite{Zhu3}]. Here below we report the numerical results for the self-collimation case. In this analysis, we considered the aluminum plate with an embedded $15\times12$ ABH square lattice in the center region. The interface of the flat area with the ABH slab was aligned along the $\Gamma-X(X')$ boundary. An $A_0$ plane wave at $f_0=2.2$ kHz (approximately $\Omega=0.0987$ corresponding to a square-like EFC) was used as excitation. The full field out-of-plane displacement is shown in Fig.\ref{Fig2}b which clearly confirms the occurrence of strong self-collimation. For clarity, we also report the corresponding wave filed in in the absence of the ABH lens (Fig.\ref{Fig2}c) which, as expected, results in a diverging beam. In more quantitative terms, without the acoustic lens the angle of aperture of the beam is approximately $\theta=37.4^{\circ}$ while in prresence of the lens the angle is approximately $\theta=0^{\circ}$.

\subsection{Experimental validation: long wavelength limit}

In order to validate the ABH-based design of the acoustic lens, we fabricated two different samples. Each sample was built embedding a different design of the acoustic lens. The two samples were design to validate the self-collimation and the bi-refraction effects. The waveguides were scaled down to a thickness of $0.16$ in ($\approx 4mm$) so to bring the frequency of the $A_0$ mode to a range convenient for excitation and measurement. The parameters of the lattice structure were modified accordingly resulting in $a = 30$ mm, radius $r = 14$ mm, taper exponent $m = 2.2$, and residual thickness $h_r = 1$ mm. These updated geometric parameters shifted the operating frequency range to $10-20$ kHz.

\begin{figure}[h!]
 \center{\includegraphics[scale=0.4]{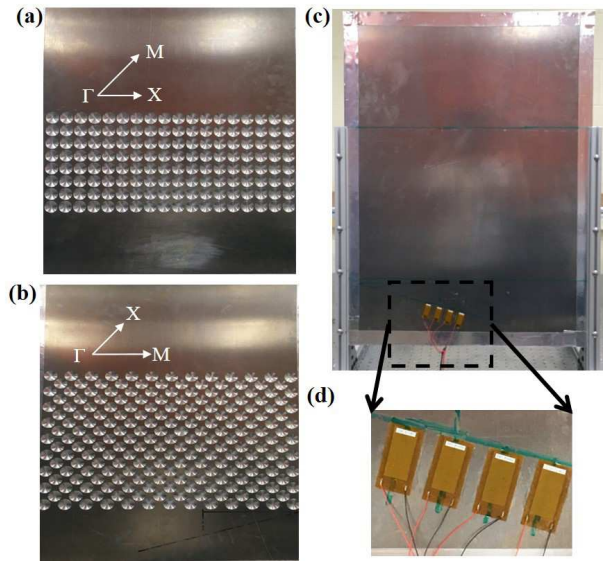}}
  \caption{Experimental setup. (a) and (b) shows the front view of the two plate samples having an embedded lens in the center section. The lenses were made out of the same ABH unit cell arranged in a square lattice structure but oriented along different directions of the Brillouin zone. In (a) the interface was aligned along the $\Gamma-X$ direction while in (b) along the $\Gamma-M$ direction. (c) The test samples were mounted in a vertical frame to facilitate the laser measurements. Viscoelastic tape was applied along the plate's edges to reduce back-scattering and reverberation. (d) The zoom-in view of the  surface mounted MFC transducers that were generating the excitation.} \label{Fig3}
\end{figure}

The experimental setup is shown in Fig.\ref{Fig3}. Figure \ref{Fig3}a and b show the front view of the two plate samples with the embedded ABH lattice. The ABH unit cell is identical in each configuration while the arrangement of the square lattice follows different directions. In Fig.\ref{Fig3}a, the interface is aligned with the $\Gamma-X$ direction while in Fig.\ref{Fig3}b it is aligned with the $\Gamma-M$ direction. During the experimental measurements, the plates were mounted in a vertical frame providing clamped boundary conditions on the left and right edges (Fig.\ref{Fig3}c). The response of the plate was acquired using a scanning laser vibrometer which measured the out-of-plane displacement field of the entire plate. An array of four Micro Fiber Composite (MFC) patches was surface bonded and used to generate the excitation. For the bi-refraction test, the MFC array was oriented at $15^{\circ}$ as shown in the zoom-in view in Fig.\ref{Fig3}d) while for the collimation test the it was aligned parallel to the interface. The MFC patches were simultaneously actuated to generate the desired $A_0$ planar wave fronts. Viscous damping tapes were applied along the edges to reduce the effect of back-scattering and reverberation.

\begin{figure}[h!]
	\center{\includegraphics[scale=0.42]{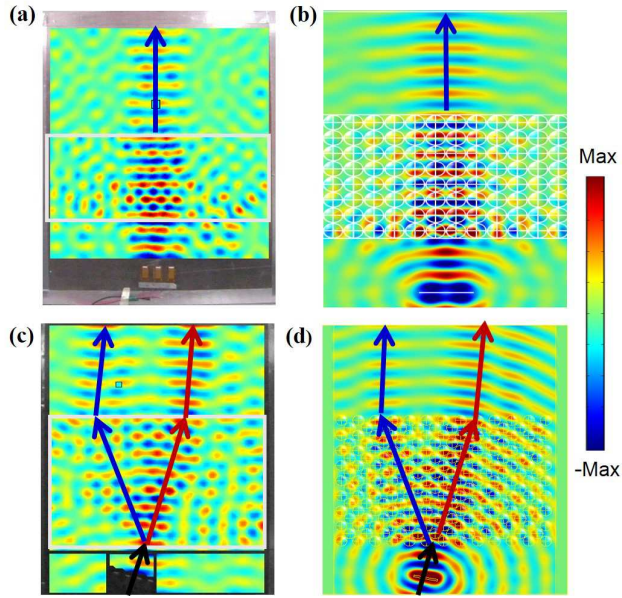}}
	\caption{Experimental results. (a) and (b) show the experimental and numerical results for the collimation design obtained at $f_1$=20.2 kHz. (c) and (d) show the experimental and numerical results for the bi-refraction case at $f_2$=13.6 kHz. In both cases, good agreement between the numerical predictions and the experimental results was observed. } \label{Fig4}
\end{figure}

The experimental results are shown in Fig.\ref{Fig4}(a),(c) and compared with the corresponding numerical simulations (Fig.\ref{Fig4}(b),(d)). The top row shows the results for collimation while the bottom row shows the bi-refraction. The area marked by the white box indicates the location of the ABH periodic lattice. Overall, there is an excellent qualitative agreement between the predicted propagation behavior and the experimental measurements. The collimation case (operating frequency $f_1=$ 20.2 kHz) shows an approximately zero angle of divergence as well as a strong persistence of the beam shape after the lens. For the bi-refraction case (operating frequency $f_2=13.6$ kHz), the incident beam is split into two beams within the lattice which give rise to a positively (red arrow) and a negatively (blue arrow) refracted beams after the lens.
Both the beams are oriented at about $5^{\circ}$ with respect to the interface, which is also in agreement with the numerical calculations.

\section{High Frequency Behavior: geometric acoustic analysis}

In this section, we analyze the performance of the lens in the high frequency range that is where the wavelength of the flexural wave is smaller than the size of the unit. In general, the response of a periodic material that operates in this range (the so-called phononic range) is dominated by scattering effects. In the case of the ABH material (as far as the smoothness condition is satisfied), flexural waves would still propagate through the lattice according to a mechanism similar to that observed in the long wavelength limit. This means that the scattering is minimized and the wave direction is gradually bent due to the spatial gradient in the phase velocity. When the wavelength is such that the smoothness condition is no longer satisfied than the response of the ABH material would be purely dominated by scattering effects, that is the classical phononic range. This suggest that, for an ABH lattice, the occurrence of the phonic range is not dominated by the size of the unit cell but instead from the smoothness condition.

To study the wave propagation through the ABH material in the high frequency range, we use both a geometric acoustic and a full field simulation approach. Geometric acoustics was chosen because it allows a good qualitative understanding of the wave behavior inside the lens due to its ability to track the ray trajectories through local inhomogeneities. 

The trajectory of a ray for uncoupled elastic waves is given by\cite{ray}:

\begin{subequations}
\begin{gather}
              \frac{d\vec{x}}{dt}=c^2\vec{s} \label{ray1} \\
              \frac{d\vec{s}}{dt}=-\frac{1}{c}\nabla c \label{ray2}
\end{gather}
\end{subequations}

where $\vec{x}$ is the position vector along the ray trajectory, $\vec{s}$ is the wave slowness vector that can be expressed as $\vec{s}=\frac{\vec{n}}{c}$ being $c$ the local phase velocity and $\vec{n}$ the unit vector normal of the wavefront. Equations \ref{ray1}-\ref{ray2} can be numerically integrated to obtain the ray trajectories across the inhomogeneous medium. For the ABH, the spatial distribution of the phase velocity is given by $\frac{c}{c_0}=\sqrt{\frac{h(r)}{h_0)}}=\sqrt{\frac{\varepsilon r^m+h_r}{h_0}}$, where $c_0$ is the phase velocity in the background material. 

\begin{figure}[h!]
	\center{\includegraphics[scale=0.43]{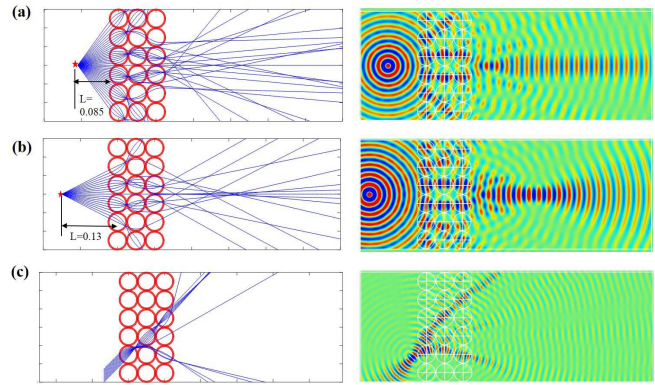}}
	\caption{Different acoustic lens effects can be achieved via periodic ABH tapers also in the short wavelength limit: including (a) collimation (b) focusing and (c) bi-refraction. The left panels show the geometric acoustic simulations while the right panels show the full wave field obtained from finite element simulations. In (a) and (b), a point source excitation at frequency $f=43$ kHz located at different distance from the lens leads to either collimation or focusing. In (c), a plane source at $f=103$ kHz and tilted at $48^{\circ}$ produces a bi-refraction effect. } \label{Fig5}
\end{figure}

The local geometric inhomogeneity of an ABH taper produces a spatial gradient in the phase velocity which bends propagating waves towards the regions with lower wave speed (i.e. the ABH center in this specific case). Generally, the larger the gradient the smaller the radius of curvature. To analyze the lens performance in the high frequency limit, we considered a flat plate with an embedded slab of ABH tapers. The slab consisted of a $3\times6$ periodic square lattice with constant $a_1=0.05$ m (Fig.\ref{Fig1}a). The ABHs had residual thickness $h_r=1.72$  mm, taper coefficient $m=2.2$, and the radius $r=24$ mm.

 Following the analyses in the long wavelength domain, we investigated the response of the plate with respect to collimation, and bi-refraction. For completeness, we also added the case of high frequency focusing. The results are summarized in Fig.\ref{Fig5} (a), (b) and (c). The left panels show the ray acoustic results while the right panels show the full field finite element simulations. 
 
 Fig.\ref{Fig5}a illustrates the collimation effects where the ABH slab is excited by a point source (see red star) at a distance $L=0.085$ m from the ABH. Aside from some lateral scattering resulting from rays at large angle of incidence, most of the rays are effectively bent and molded into a collimated beam after the lens. The ray acoustics simulations clearly show the operating mechanism within the lens and confirm that, also in the high frequency regime, the ABHs induce a progressive bending of the rays instead of an abrupt change in the propagation direction typical of traditional high-frequency scattering and refraction. The full wave simulation analysis at $f_0$=43 kHz (Fig.\ref{Fig5}a, right panel), that is corresponding to a wavelength of $\lambda =0.598a_1$  also shows the formation of a collimated beam that is fully consistent with the ray acoustic predictions. For completeness, we show also the high-frequency focusing effect from the same point source that can be achieved at the same frequency by adjusting the position of the source ($L=0.13$ m) with respect to the lens (Fig.\ref{Fig5}b). Both ray acoustics and full field results are fully consistent with each other and show that the focusing effective is still due to a progressive bending of the rays.

Finally, figures \ref{Fig5}c show the predictions for the high-frequency bi-refraction case. A planar wave front at an angle of $48^{\circ}$ was used to illuminate the lens. In the geometric acoustics analysis, the source was simulated as a beam of initially parallel rays. The simulations show that the original beam is separated into two beams that are progressively bent in diverging directions, therefore giving rise to positive and negative refraction. Once again, these results show that the operating mechanism is unchanged with respect to the long wavelength regime. We merely note that, similarly to what found for the metamaterial range, the bi-refraction effect will occur only above a given value of the incident angle. This suggests the existence of a critical angle analogously to classical refraction phenomena. This propagation mechanism was further confirmed by full field analysis whose results are shown in Fig. \ref{Fig5}c at $f_0=103$ kHz,that is corresponding to a wavelength of $\lambda =0.3267a_1$ .

\begin{figure}[h!]
	\center{\includegraphics[scale=0.44]{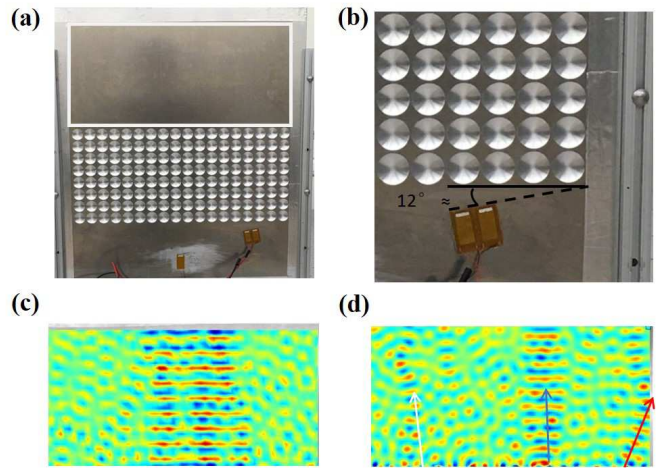}}
	\caption{(a) Front view of the test bed. The white box indicates the area where measurements are performed. (b) Zoomed-in view shows the MFC array tilted at an angle of $12^{\circ}$ with respect to the ABH lattice. (c) and (d) show the experimentally measured displacement fields for the case of collimation and bi-refraction, respectively.} \label{Fig6}
\end{figure}

A simplified experimental study was performed to validate the above analyses. The experimental setup retained for the test was the same used for the long wavelength limit experimental validation. We chose to maintain the same experimental setup in order to show the ability of the same lens to operate in the two regimes. Nevertheless, note that to achieve optimal performance across different operating regimes the lens would still require some tuning. As an example, in the short wavelength limit this tuning would lead towards a lower number of ABH rows, which is consistent with the design approach followed in the numerical simulations. Both the collimation and bi-refraction cases were explored (Fig. \ref{Fig6}). The actuation in these two cases was performed using: 1) a single MFC acting as a quasi-point-source, and 2) an array of two MFCs at an oblique incidence of approximately $12^{\circ}$ (Fig.\ref{Fig6}b). The white box in Fig. \ref{Fig6}a indicates the area after the ABH lattice where measurements were performed. The lattice was excited harmonically at a frequency  $f=50$ kHz (corresponding to a wavelength of $\lambda =0.8167a$) and the response was measured using a scanning laser vibrometer. The experimental results are shown in Fig. \ref{Fig6}c and d which correspond to the point source and to the oblique incidence excitation, respectively. We must highlight that the frequency of excitation was right on the bandwidth limit of the transducer amplifier which limited considerably the maximum obtainable signal-to-noise ratio. Despite this hardware limitation, in both cases results were sufficiently clear to show the occurrence of both collimation and bi-refraction, as predicted. In Fig. \ref{Fig6}d a second positively refracted beam (marked by the white arrow), is also visible due to the large width of the lens as well as refractions from the boundaries. As mentioned above, for optimal results the lens parameters should be adjusted. 

We believe that these results supports our initial statement that the ABH-based design enables the synthesis of acoustic metamaterials characterized by a broad operating range. This result is possible because both regimes exploit, within the limit of the smoothness condition, the same fundamental operating principle that is a progressive bending of the acoustic ray due to a phase velocity gradient. This characteristic is in net contrast with the more traditional designs of engineered materials that are typically targeted to either the metamaterial or the phononic range.

\section{Conclusion}

We investigated the use of geometric inhomogeneities in order to create broadband acoustic lenses that can be fully embedded in thin-walled structural waveguides without altering their basic structural characteristics. The study was also targeted to demonstrate that these materials operate in the two wavelength regimes according to the same mechanism, that is a progressive bending of the wave trajectory due to a spatially tailored wave speed profile. In fact, it is this mechanism that allows such materials to achieve broadband performance over both the metamaterial and the phononic range (within the the validity limit of the smoothness assumption). The specific type of inhomogeneity used in this study, known as the Acoustic Black Hole (ABH), consists in a geometric taper that follows a power-law profile. Numerical simulations performed via both geometric acoustics and finite element analysis have shown the ability of these materials to achieve collimation, bi-refraction, and focusing in both the long and short wavelength limits. An experimental investigation was performed to validate the design approach and the performance in both wavelength regimes. The experimental results were in very good qualitative agreement with the numerical predictions therefore confirming the validity of the models used to understand the fundamental operating principles of ABH materials.

\section{Acknowledgments}
The authors gratefully acknowledge the financial support of the Air Force Office of Scientific Research under the YIP grant \#FA9550-15-1-0133.

\bibliographystyle{apsrev}
\bibliography{ABH_Lens_ref}

\end{document}